\documentclass[letterpaper]{article} 
\usepackage[]{anonymous}  
\usepackage{times}  
\usepackage{helvet}  
\usepackage{courier}  
\usepackage[hyphens]{url}  
\usepackage{graphicx} 
\usepackage{natbib}  
\usepackage{caption} 
\usepackage{algorithm}
\usepackage{algorithmic}
\usepackage{amsmath} 
\usepackage{newfloat}
\usepackage{listings}
\usepackage{nameref}
\usepackage{multirow}
\usepackage{amssymb}
\usepackage{mathtools}
\usepackage{amsthm}
\usepackage{multirow}
\usepackage{twemojis}
\usepackage{makecell}
\usepackage{pifont}
\DeclareCaptionStyle{ruled}{labelfont=normalfont,labelsep=colon,strut=off} 
\floatstyle{ruled}
\newfloat{listing}{tb}{lst}{}
\floatname{listing}{Listing}
\title{Lossless Copyright Protection via Intrinsic Model Fingerprinting}
\author {
	Lingxiao Chen\textsuperscript{\rm 1},
    Liqin Wang\textsuperscript{\rm 1},
	Wei Lu\textsuperscript{\rm 1}\thanks{Corresponding authors},
    Xiangyang Luo\textsuperscript{\rm 2}
}
\affiliations {
	\textsuperscript{\rm 1}MoE Key Laboratory of Information Technology, Sun Yat-sen University, Guangzhou, China\\
    \textsuperscript{\rm 2}State Key Laboratory of Mathematical Engineering and Advanced Computing, Zhengzhou, China\\
	chenlx67@mail2.sysu.edu.cn, wanglq37@mail2.sysu.edu.cn, luwei3@mail.sysu.edu.cn, luoxy\_ieu@sina.com
}

\usepackage{bibentry}

\begin{document}

\maketitle

\begin{abstract}

The exceptional performance of diffusion models establishes them as high-value intellectual property but exposes them to unauthorized replication. 
Existing protection methods either modify the model to embed watermarks, which impairs performance, or extract model fingerprints by manipulating the denoising process, rendering them incompatible with black-box APIs. 
In this paper, we propose TrajPrint, a completely lossless and training-free framework that verifies model copyright by extracting unique manifold fingerprints formed during deterministic generation. 
Specifically, we first utilize a watermarked image as an anchor and exactly trace the path back to its trajectory origin, effectively locking the model fingerprint mapped by this path. Subsequently, we implement a joint optimization strategy that employs dual-end anchoring to synthesize a specific fingerprint noise, which strictly adheres to the target manifold for robust watermark recovery. As input, it enables the protected target model to recover the watermarked image, while failing on non-target models. Finally, we achieved verification via atomic inference and statistical hypothesis testing.
Extensive experiments demonstrate that TrajPrint achieves lossless verification in black-box API scenarios with superior robustness against model modifications.

\end{abstract}


\section{Introduction}
The versatility and high fidelity of diffusion models \cite{diffusion2015,ddpm} in image synthesis, editing, and multimodal generation \cite{diffusion_survey} have rendered them integral to the field of generative artificial intelligence. The training of high-quality models such as Stable Diffusion \cite{ldm} and DALL-E \cite{ramesh2021zero} necessitates substantial computational resources and vast datasets, establishing them as valuable intellectual property. However, the open-source nature of model weights and the proliferation of model distillation techniques \cite{salimans2022} have introduced significant vulnerabilities. These factors render proprietary models highly susceptible to misappropriation, illegal replication, and unauthorized usage \cite{jiang2023comprehensive,tramer2016stealing}. Consequently, securing model copyright without compromising utility remains a critical challenge.

Existing copyright protection strategies employ active intervention, categorized into inference intervention and model fine-tuning. Inference-based methods \cite{treering,gaussianshading,robin,shallowdiffuse} inject specific signals into the initial noise or the denoising loop. However, forcing such injection disrupts natural diffusion trajectories, which introduces artifacts and limits diversity \cite{fingerinv}. In contrast, fine-tuning approaches \cite{stable_signature,AquaLoRA,sleepermark} implant watermarks by modifying model parameters. This increases training costs and deviates from the pre-trained optimal distribution, thereby impairing generation performance \cite{watermarkvisualconcepts}. To avoid such degradation, model fingerprinting offers a lossless alternative by extracting inherent behavioral features without modifying parameters.

\begin{figure}
\centering
{\includegraphics[width=1.0\linewidth]{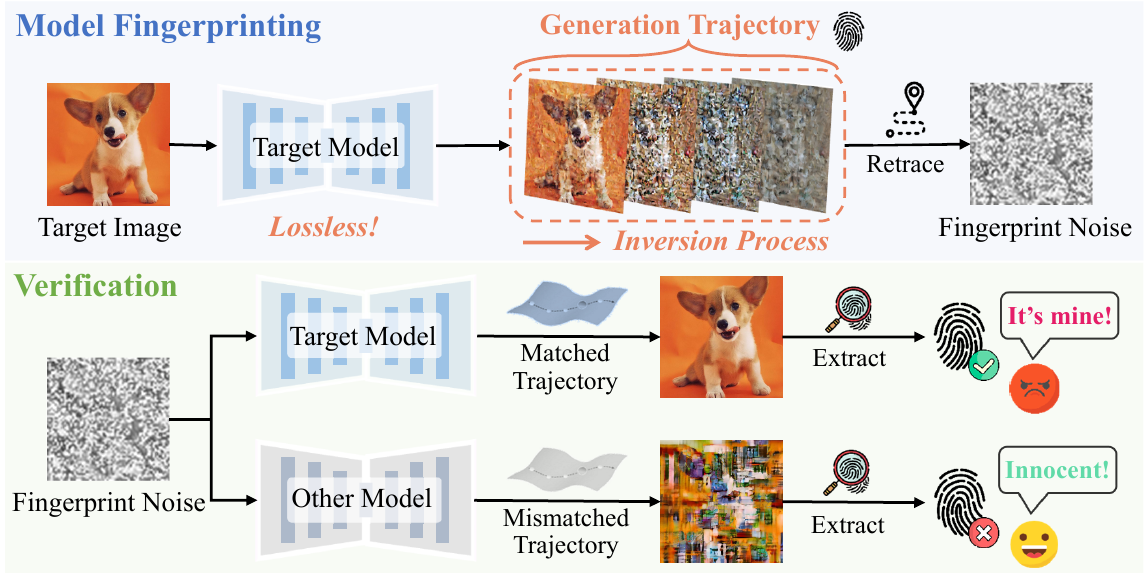}}
\caption{Illustration of the TrajPrint. We retrace the generative trajectory to bind a specific fingerprint noise to the model's intrinsic path. This noise acts as a fingerprint carrier, triggering valid watermark reconstruction on the target model due to manifold match while remaining invalid on other models.}
\label{introduction}
\end{figure}

Despite the significant potential of model fingerprinting and its proven effectiveness in discriminative models \cite{Fingerprinting1,Fingerprinting2,Fingerprinting3}, applying this technology to diffusion models faces implementation hurdles. Existing extensions typically require fine-grained intervention in intermediate denoising steps \cite{fingerinv}, rendering them incompatible with black-box APIs restricted to atomic inference, where generation is an indivisible operation without intermediate access. Furthermore, these methods lack rigorous statistical criteria, thereby failing to provide quantifiable evidence for reliable verification.

To address these challenges, we propose TrajPrint for model copyright verification. As shown in Figure \ref{introduction}, our method is completely lossless, training-free, and compatible with black-box settings. We posit that the deterministic generative process of a diffusion model constitutes a unique high-dimensional manifold mapping. This intrinsic mapping serves as the model's unique identity, manifesting as specific denoising trajectories. Based on this insight, we regard the endogenous generative trajectory as a behavioral fingerprint that distinguishes the target model from others. We retrace this trajectory to identify a specific input noise that binds to the intrinsic generative path. This specific noise successfully triggers the reconstruction of the target image on the target model, whereas it remains inert on non-target models due to manifold mismatch. Consequently, this noise functions as the carrier for the model fingerprint. By mapping the generative manifold to the fingerprint noise, the model owner obtains a unique trigger-verification pair, enabling credible copyright verification without compromising the original model generation quality.

Specifically, we first employ trajectory inversion to lock the origin of the model's denoising path, which serves as the preliminary fingerprint noise. Then we construct a watermarked anchor containing embedded copyright messages to serve as the fixed trajectory endpoint. We map this anchor back into the latent noise space to locate the unique trajectory origin using deterministic inversion. This step effectively anchors the verification process guided by the fingerprint noise to the specific denoising path determined by the model weights.
Subsequently, to mitigate the impact of discretization errors inherent in inversion, we propose a joint optimization strategy to synthesize the final fingerprint noise. We employ a trajectory anchoring mechanism that enforces perceptual manifold alignment at the output while constraining the optimized noise to the neighborhood of the trajectory origin. This ensures the optimized noise specifically activates only the target model’s specific manifold while preventing the generation of universal adversarial perturbations.
Finally, we feed the optimized fingerprint noise into a suspect model and determine copyright ownership by calculating the bit accuracy of the extracted watermark. We quantify verification confidence using a one-sample t-test, providing statistically significant evidence of infringement. Our contributions are as follows:
\begin{itemize}
\item We propose a completely lossless and training-free method for model copyright verification. By extracting intrinsic model fingerprints, our approach ensures zero degradation of generation quality and remains compatible with encapsulated black-box APIs.
\item We map the unique generative manifold to a specific fingerprint noise by retracing the model's intrinsic denoising trajectory, thereby constructing a trigger-verification pair to achieve lossless model verification.
\item Extensive experiments demonstrate that TrajPrint achieves reliable copyright verification across diffusion models with diverse architectures and exhibits robustness against model modification attacks.
\end{itemize}

\section{Related Works}

\subsection{Intellectual Property Protection for Diffusion Models}
Intellectual property protection for diffusion models primarily relies on active intervention. These strategies are generally categorized into inference intervention and model fine-tuning. Inference intervention methods inject watermarks during generation without weight modification. Approaches like Tree-Ring \cite{treering} and Gaussian Shading \cite{gaussianshading} target the initialization phase, while RoBin \cite{robin} and Shallow Diffuse \cite{shallowdiffuse} constrain intermediate latent variables. However, such forced interventions often disrupt natural denoising trajectories, limiting generation diversity.
Model fine-tuning methods modify parameters to enhance robustness. Stable Signature \cite{stable_signature} fine-tunes the decoder, while AquaLoRA \cite{AquaLoRA} and SleeperMark \cite{sleepermark} embed signals directly into the U-Net. However, altering pre-trained weights deviates from the optimal distribution, degrading generation quality.

Distinct from active intervention, model fingerprinting offers a lossless alternative by extracting intrinsic behaviors. However, approaches like FingerInv \cite{fingerinv} mandate fine-grained sampling intervention, which preclude atomic inference in black-box APIs and lack rigorous statistical criteria for reliable verification.

\begin{figure*}
\centering
{\includegraphics[width=0.85\linewidth]{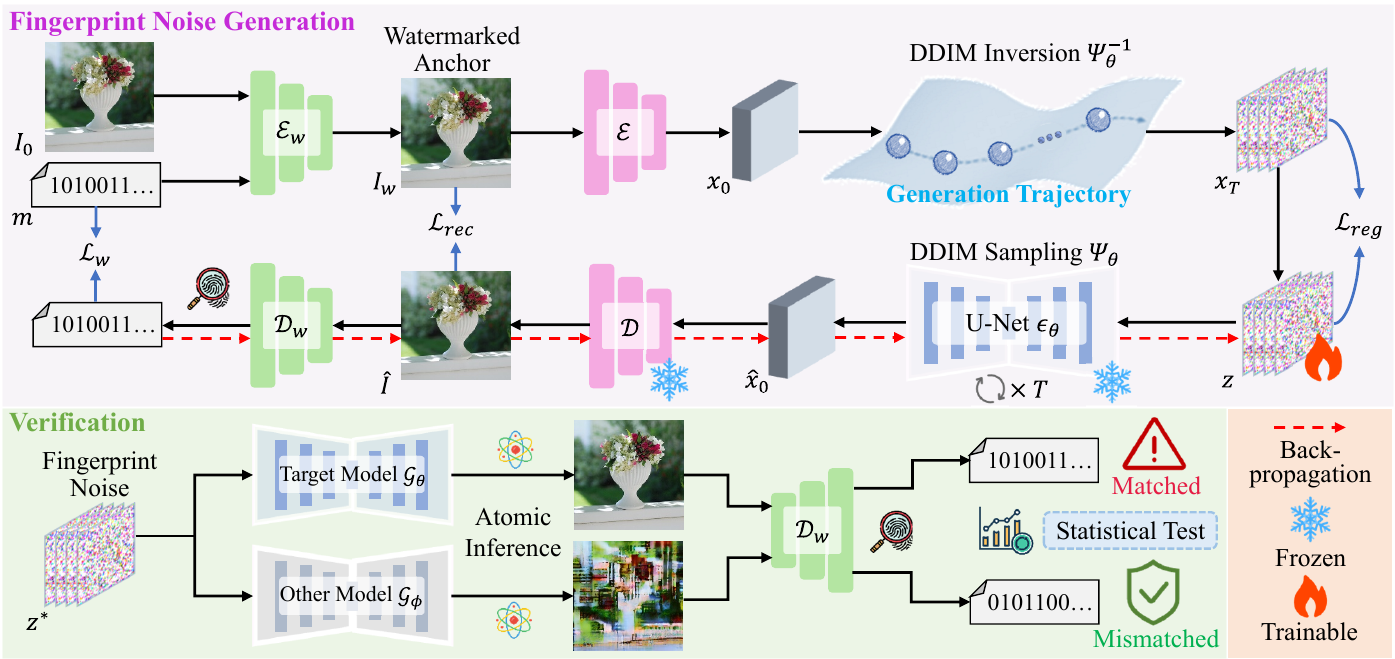}}
\caption{Overview of the proposed TrajPrint. (1) Fingerprint noise generation: We construct a watermarked anchor containing a binary message. Initialized via DDIM inversion, the fingerprint noise is optimized end-to-end through the frozen diffusion model under dual-end constraints, ensuring its generation trajectory targets the anchor. (2) Verification: The optimized noise is fed into candidate models. We decode the watermark from the generated images and apply a statistical test to achieve high-confidence copyright verification.}
\label{overview}
\end{figure*}

\subsection{Diffusion Models and DDIM Inversion}
Diffusion models are fundamentally conceptualized as dual processes \cite{ldm}. The forward process progressively perturbs clean data into Gaussian noise, while the reverse process recovers the original data using a learned denoising network to approximate the score function. Although standard diffusion models often employ stochastic sampling \cite{ddpm}, denoising diffusion implicit models (DDIM) \cite{ddim} characterize the process as a deterministic probability flow ordinary differential equation (ODE). In this framework, sampling is treated as the discretized solution to the ODE, establishing a strictly deterministic mapping from the initial latent noise to the final image. Crucially, these ODE trajectories ensure bi-directional consistency and exhibit mathematically reversibility. This allows for integration along the temporal axis in reverse, enabling deterministic inversion from the image domain back to the noise domain \cite{ddiminversion}. While such inversion is commonly applied in image editing tasks \cite{ddimedit1,ddimedit2}, our work creatively repurposes it as a black-box security tool to extract inherent model fingerprints.

\section{Problem Statement}

\subsection{Threat Model}
\textbf{Attacker Capabilities.}
In common scenarios of model copyright protection, the attacker illicitly obtains the proprietary source model from the owner and deploys it as an independent service. To evade copyright detection or adapt the model for downstream requirements, the attacker may apply various model modifications. Specifically, they may employ parameter-efficient fine-tuning techniques, such as LoRA \cite{lora} and DreamBooth \cite{dreambooth}, to alter the generative style. Furthermore, they might implement structural changes like model quantization and pruning to reduce deployment costs.

\textbf{Defender Capabilities.}
The defender holds the copyright to the source model and maintains full white-box access for offline preparation. Conversely, during the verification phase against a suspect model, the defender operates under strict black-box constraints. Access is limited to querying the public interface without visibility into internal parameters, gradients, or intermediate latent states. The verification is restricted to atomic inference, lacking the ability to observe or manipulate intermediate states.

\subsection{Design Goals}
To achieve copyright verification under the defined threat model, an ideal fingerprinting scheme must satisfy three core properties:
(i) \textbf{Effectiveness}. The fingerprint must be reliably identified by the target model, triggering a definitive and scientifically credible response to establish ownership. We must ensure the verification signal remains robust even after attackers apply modifications like fine-tuning or quantization.
(ii) \textbf{Specificity}. The fingerprint must remain invalid on non-target models to prevent false accusations. We must ensure the fingerprint is anchored to the unique intrinsic manifold of the target model, rather than relying on generic transferable noise.
(iii) \textbf{Losslessness}. Fingerprint extraction should be performed without modifying the model's original weights or architecture, as we need to prevent degradation of the original model generation quality.

\section{Method}

\subsection{Overview}
Our framework is illustrated in Figure \ref{overview}.
To ensure losslessness, we aim to construct a specific fingerprint noise $z^*$ that acts as a unique trigger for the target model $G_\theta$. We first construct a watermarked anchor, which is an image embedded with a discrete copyright message. Using deterministic DDIM inversion, we map this anchor back to its generation trajectory noise origin in the latent space, effectively locking the model's intrinsic generative behavior. To ensure effectiveness against inversion errors and enhance specificity, we then perform a joint optimization to refine this noise. The optimization constrains the generation process at dual ends: it enforces perceptual manifold alignment at the output while regularizing the optimized fingerprint noise $z^*$ to remain in the vicinity of the model-specific origin noise $x_{T}$.
In the verification phase, the defender performs atomic inference on the suspect model using the fingerprint noise $z^*$. Due to manifold matching, only the protected target model successfully reproduces the hidden watermark. Subsequently, the defender verifies the copyright by calculating the bit accuracy and conducting a one-sample t-test to provide statistical evidence of effectiveness.

\subsection{Trajectory Inversion}
\label{trajectory_inversion}
Theoretically, the generative process of diffusion models can be formulated as the discretized solution to a deterministic probability flow ordinary differential equation (PF-ODE). This theoretical framework is practically realized by denoising diffusion implicit models (DDIM), which redefines the generation as a deterministic process. We denote the DDIM sampler as a bijective mapping $\Psi_\theta: \mathcal{X}_T \to \mathcal{X}_0$ operating within the latent space, where $\mathcal{X}_T$ represents the initial noise distribution governed by $\mathcal{N}(\mathbf{0}, \mathbf{I})$ and $\mathcal{X}_0$ denotes the latent data distribution. For any given initial noise $x_T \in \mathcal{X}_T$, the mapping ${x}_0 = \Psi_\theta({x}_T)$ is strictly determined by the model parameters $\theta$. This bijective property of DDIM implies that $\Psi_\theta$ defines a unique high-dimensional manifold mapping, serving as a natural, intrinsic model fingerprint. Our goal is to extract this fingerprint by exploiting the invertibility of the DDIM sampler, ensuring a completely lossless process.

However, practically verifying this fingerprint requires a quantifiable standard. Relying on visual similarity by directly reconstructing images from noise to determine infringement lacks absolute certainty. Therefore, to effectively extract fingerprints and establish a statistical standard for credible verification, we transform verification into a watermark recovery task. Specifically, we embed a binary copyright message $m \in \{0, 1\}^k$ into a carrier image $I_0$ using a pre-trained encoder $\mathcal{E}_w$, yielding a watermarked anchor $I_w$.
This watermarked anchor acts as a fixed boundary condition at timestep $t=0$, binding the discrete copyright signal $m$ to a specific coordinate on the model's manifold.

\begin{figure}
\centering
{\includegraphics[width=0.95\linewidth]{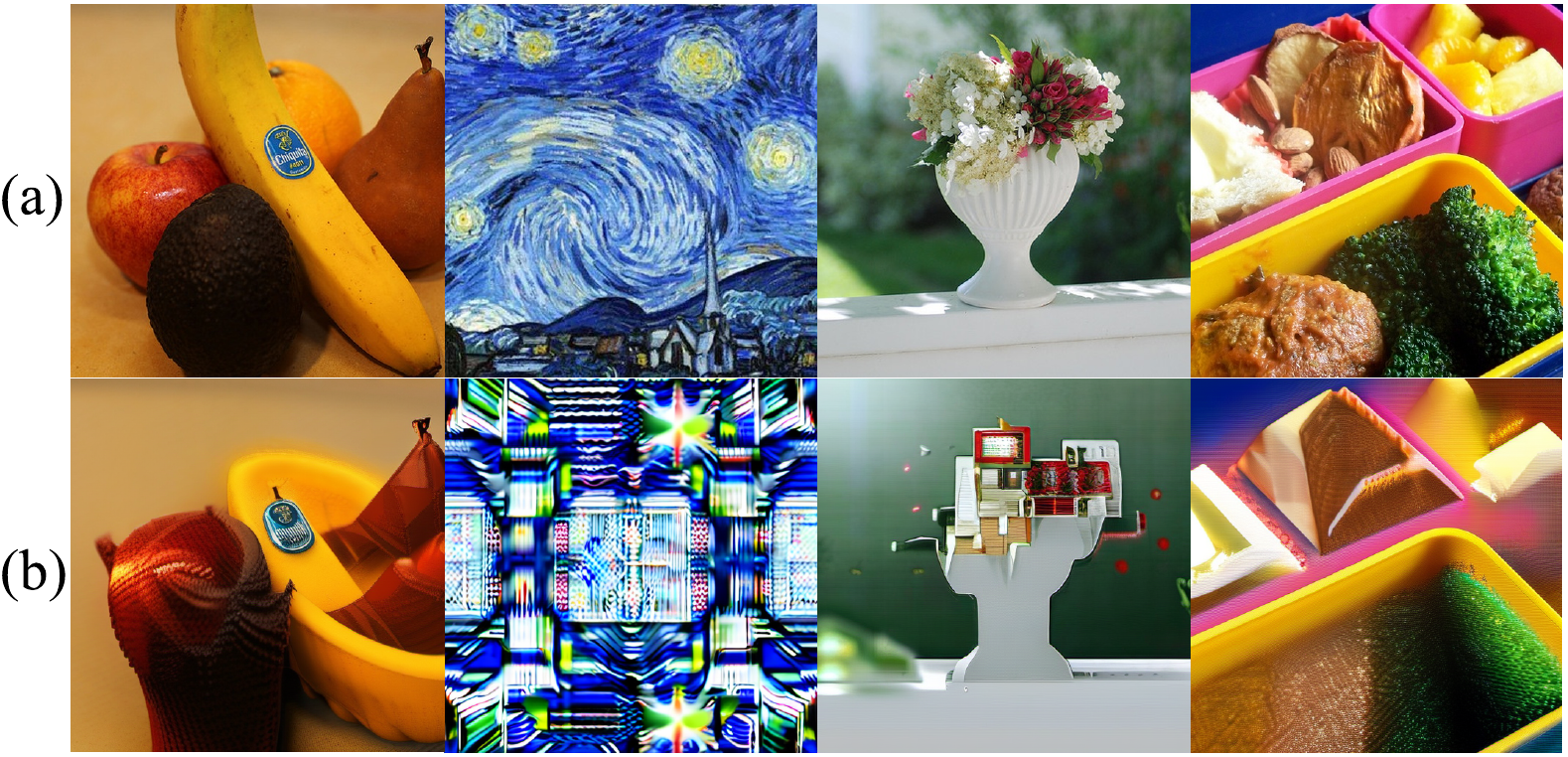}}
\caption{Visualization of reconstruction errors in direct DDIM inversion. (a) Original anchor images. (b) Reconstructed images.}
\label{method}
\end{figure}

To trace the unique generation path pointing to this anchor, we employ DDIM inversion, the exact reverse process of the sampler, denoted as $\Psi^{-1}_\theta: \mathcal{X}_0 \to \mathcal{X}_T$. We first project the anchor into the latent space via the VAE encoder, $x_0 = \mathcal{E}(I_w)$, and then execute the DDIM inversion process (from $t=0$ to $t=T$) to derive the terminal noise state:
\begin{equation}
x_T = \Psi^{-1}_\theta(x_0)
\end{equation}
Here, $x_T$ represents the unique projection of the watermarked anchor onto the latent manifold of $G_\theta$. However, direct reconstruction from $x_T$ often leads to visual distortions and the loss of high-frequency details due to discretization errors inherent in the inversion process, ultimately resulting in the destruction of watermark information, as illustrated in Figure \ref{method}.
To compensate for these inversion errors, we introduce an optimizable latent variable $z$, which is initialized with the trajectory origin $x_T$. We then define the watermark recovery loss to guide the optimization of $z$:
\begin{equation}
L_w(z) = \text{BCE}(\mathcal{D}_w(\mathcal{D}(\Psi_\theta(z))), m)
\end{equation}
where $\mathcal{D}$ denotes the image decoder and $\mathcal{D}_w$ is the differentiable watermark extraction network. This objective $L_w$ guides the optimization of $z$, ensuring that the generated image effectively recovers the embedded message $m$.

\subsection{Trajectory Anchoring}
To guarantee high fingerprint specificity, we introduce a trajectory anchoring mechanism. It imposes dual-end constraints during optimization, strictly confining the generation trajectory to the target model's intrinsic manifold.

\textbf{Perceptual Manifold Alignment.} We first anchor the trajectory at the output end by enforcing semantic consistency. By fixing the generation endpoint to $I_w$, we constrain the ODE trajectory to converge to a specific location on the image manifold. This compels the optimization to discover the input noise that aligns with the target model's unique denoising path.
Therefore, we define a visual reconstruction loss that compels the generated image to visually align with the watermarked anchor $I_w$:
\begin{equation}
L_{rec}(z) = \|D(\Psi_\theta(z)) - I_w\|_2^2 + \gamma L_{p}(D(\Psi_\theta(z)), I_w)
\end{equation}
where $L_{p}$ denotes the perceptual loss \cite{wastonloss} and $\gamma$ is a weighting factor. This term serves as a semantic constraint that anchors the noise to the target model. Since distinct diffusion models map the same noise to divergent semantic outputs, enforcing reconstruction ensures that the noise $z^*$ effectively exploits the unique parameter weights of $\mathcal{G}_\theta$. If an unrelated model $\mathcal{G}_\phi$ processes $z^*$, the resulting image will deviate significantly from $I_w$, destroying the spatial structure required for watermark decoding. Thus, $L_{rec}$ effectively couples the watermark recoverability with the target model's specific semantic generation capabilities.

\textbf{Latent Trajectory Regularization.} Complementarily, we anchor the trajectory at the input end. We constrain the optimized noise $z^*$ to remain within the local neighborhood of the model-specific trajectory origin $x_T$:
\begin{equation}
L_{reg}(z) = \|z - x_T\|_2^2
\end{equation}
This regularization restricts the search space to the vicinity of the intrinsic trajectory. It geometrically confines the search space, precluding divergence into regions harboring universal adversarial examples. By constraining the fingerprint to $x_T$, we ensure the optimization remains a minimal perturbation of the model's natural behavior rather than an off-manifold adversarial perturbation.

\begin{algorithm}[tb]
   \caption{The Algorithm of TrajPrint}
   \label{Algorithm}
\begin{algorithmic}[1]
   \STATE {\bfseries Input:} Diffusion denoiser $\epsilon_\theta$, VAE $\mathcal{E}/\mathcal{D}$, watermark encoder and decoder $\mathcal{E}_{w}/\mathcal{D}_{w}$, image $I_{0}$, message $m$, timesteps $T$, iterations $N$, learning rate $\eta$
   \STATE {\bfseries Output:} Optimized fingerprint noise $z^*$
   
   \STATE \textcolor{blue}{\textit{// Trajectory Inversion}}
   \STATE $I_{w} = \mathcal{E}_{w}(I_{0}, m)$, $x_0 = \mathcal{E}(I_{w})$
   \STATE $x_{T} = \Psi_{\theta}^{-1}(x_{0})$, $z = x_{T}$
   \FOR{$n=1$ {\bfseries to} $N$}
       \STATE $\hat{x}_0 = \Psi_{\theta}(z)$, $\hat{I} = \mathcal{D}(\hat{x}_0)$
       \STATE $\mathcal{L}_w(z) = \text{BCE}(\mathcal{D}_{w}(\hat{I}), m)$
       \STATE \textcolor{blue}{\textit{// Trajectory Anchoring}}
       \STATE $\mathcal{L}_{rec}(z) = \| \hat{I} - I_w \|_2^2 + \gamma \mathcal{L}_{p}(\hat{I} - I_w)$
       \STATE $\mathcal{L}_{reg}(z) = \| z - x_{T} \|_2^2$
       \STATE \textcolor{blue}{\textit{// Joint Optimization}}
       \STATE $\mathcal{L}_{total}(z) = \mathcal{L}_{w}(z) + \lambda_{rec}\mathcal{L}_{rec}(z) + \lambda_{reg}\mathcal{L}_{reg}(z)$
       \STATE $z = z - \eta \cdot \nabla_z \mathcal{L}_{total}$
   \ENDFOR 
   \STATE \textbf{return} $z^*=z$
\end{algorithmic}
\end{algorithm}

\textbf{Joint Optimizing Objective.} We integrate the aforementioned objectives into a unified optimization framework. The total loss function is defined as the weighted sum of the watermark recovery objective and the trajectory anchoring constraints:
\begin{equation}
L_{total}(z) = L_w(z) + \lambda_{rec}L_{rec}(z) + \lambda_{reg}L_{reg}(z)
\end{equation}
where $\lambda_{rec}$ and $\lambda_{reg}$ are hyperparameters governing the trade-off between semantic alignment and latent regularization, respectively. The optimization is performed exclusively on the latent input $z$, initialized from the trajectory origin $x_T$, while keeping the model parameters $\theta$ frozen. Following \cite{attackresilient}, we employ gradient checkpointing to enable efficient end-to-end optimization. The complete TrajPrint procedure is detailed in Algorithm \ref{Algorithm}.

\subsection{Verification}
During the verification phase, the defender queries the suspect black-box model $\mathcal{G}_\phi$ with the optimized fingerprint noise $z^*$. We extract the message $\hat{m}$ from the generated image $\hat{I} = \mathcal{G}_\phi(z^*)$ using the decoder $D_w$ and calculate the bit accuracy (BA), which is defined as:
\begin{equation}
\text{BA}_i = 1 - \frac{1}{L} \sum_{j=1}^{L} \left| m_{i,j} - \hat{m}_{i,j} \right|
\end{equation}
To provide legally robust evidence, we formulate the decision process as a one-sample t-test. We define the Null Hypothesis ($H_0$) as the scenario where the suspect model is unrelated to the target, implying the watermark extraction is equivalent to random guessing ($\mu_0 = 0.5$). We calculate the t-statistic based on the sample mean $\bar{\mu}$ and standard deviation $s$ over $N$ verification samples:
\begin{equation}
t = \frac{\bar{\mu} - 0.5}{s / \sqrt{N}}
\end{equation}
If the calculated $p$-value falls below a significance level $\alpha$, we reject $H_0$ with high confidence, confirming that the suspect model $\mathcal{G}_\phi$ shares the specific fingerprint of the owner's model, thereby providing strong evidence of piracy.

\section{Experiments}

\subsection{Experimental Setup}
\textbf{Datasets and Models.} 
To evaluate architectural generalization, we employed diverse latent diffusion models: the U-Net-based Stable Diffusion v1.5, v2.1, and XL \cite{SDXL}, U-Net-NAS-based DeciDiffusion \cite{DeciDiffusion}, and DiT-based Pixart-$\alpha$ \cite{pixartalpha}. All models are executed on an NVIDIA RTX A6000 GPU using float32 precision. For verification carriers, we sample images from MS-COCO \cite{coco} and WikiArt \cite{wikiart}, covering diverse data distributions such as natural scenes and artistic paintings.

\begin{table*}[]
    \centering
    \caption{Quantitative results of cross-model verification. We compare the bit accuracy and $p$-values of our method against the random noise baseline across five different diffusion models. Columns denote the models used for fingerprint noise optimization, while rows represent the models performing verification.}
    \label{main_experiment}
    \resizebox{0.95\textwidth}{!}{
    \begin{tabular}{c l c c c c c c c c c c}
        \hline\hline
        \multirow{2}{*}{Methods} & \multirow{2}{*}{Models} & \multicolumn{2}{c}{SD1.5} & \multicolumn{2}{c}{SD2.1} & \multicolumn{2}{c}{SDXL} & \multicolumn{2}{c}{Deci} & \multicolumn{2}{c}{Pixart} \\
        \cline{3-12} 
        & & Bit Acc & $p$-value & Bit Acc & $p$-value & Bit Acc & $p$-value & Bit Acc & $p$-value & Bit Acc & $p$-value \\
        \hline
        \multirow{5}{*}{\shortstack[c]{Random\\Noise}} 
        & SD1.5           & \textbf{0.937} & \textbf{$<$1e-3} & 0.765 & $<$1e-3 & 0.708 & $<$1e-3 & 0.713 & $<$1e-3 & 0.684 & $<$1e-3 \\
        & SD2.1           & 0.729 & $<$1e-3 & \textbf{0.935} & \textbf{$<$1e-3} & 0.685 & $<$1e-3 & 0.710 & $<$1e-3 & 0.691 & $<$1e-3 \\
        & SDXL            & 0.675 & $<$1e-3 & 0.665 & $<$1e-3 & \textbf{0.885} & \textbf{$<$1e-3} & 0.678 & $<$1e-3 & 0.658 & $<$1e-3 \\
        & Deci   & 0.686 & $<$1e-3 & 0.712 & $<$1e-3 & 0.695 & $<$1e-3 & \textbf{0.862} & \textbf{$<$1e-3} & 0.689 & $<$1e-3 \\
        & Pixart & 0.667 & $<$1e-3 & 0.680 & $<$1e-3 & 0.668 & $<$1e-3 & 0.671 & $<$1e-3 & \textbf{0.932} & \textbf{$<$1e-3} \\
        \hline
        
        \multirow{5}{*}{\shortstack[c]{TrajPrint}} 
        & SD1.5           & \textbf{0.998} & \textbf{$<$1e-3} & 0.582 & 0.051 & 0.542 & 0.203 & 0.551 & 0.154 & 0.531 & 0.267 \\
        & SD2.1           & 0.594 & 0.032 & \textbf{0.979} & \textbf{$<$1e-3} & 0.565 & 0.098 & 0.535 & 0.242 & 0.548 & 0.168 \\
        & SDXL            & 0.543 & 0.254 & 0.552 & 0.149 & \textbf{0.982} & \textbf{$<$1e-3} & 0.531 & 0.337 & 0.528 & 0.362 \\
        & Deci   & 0.545 & 0.184 & 0.538 & 0.223 & 0.525 & 0.308 & \textbf{0.986} & \textbf{$<$1e-3} & 0.554 & 0.141 \\
        & Pixart & 0.532 & 0.330 & 0.541 & 0.205 & 0.545 & 0.282 & 0.558 & 0.122 & \textbf{0.978} & \textbf{$<$1e-3} \\
        \hline\hline 
    \end{tabular}
    }
\end{table*}

\textbf{Implementation Details.} 
We utilize a pre-trained HiDDeN \cite{hidden} network with a message length of $k=48$ bits. Both DDIM inversion and sampling use $T=50$, guidance scale $=1.0$, and null prompts for unconditional generation. Images are resized to $512 \times 512$. For noise optimization, we employ the Adam optimizer for $N=200$ iterations with a learning rate of $0.1$. Loss weights are set to $\gamma=0.2$, $\lambda_{rec}=0.6$, and $\lambda_{reg}=0.05$. Finally, we utilize gradient checkpointing \cite{checkpoint} to enable efficient end-to-end backpropagation on a single GPU.

\textbf{Evaluation Metrics.} 
We evaluate verification reliability using average bit accuracy, which quantifies the fidelity of the extracted watermark. A high bit accuracy score serves as direct evidence of potential infringement. We also report $p$-values from one-sample t-tests. A $p$-value below the significance level confirms that the high bit accuracy is statistically significant and not due to random chance, providing statistically significant evidence of infringement.

\textbf{Baseline.} 
As the first framework enabling verification via atomic black-box inference, we establish an internal baseline: random noise. This method initiates optimization from standard Gaussian noise, serving as a trajectory-agnostic control. We also compare against representative state-of-the-art methods: the fingerprinting method FingerInv \cite{fingerinv}, the semantic watermarking Gaussian Shading \cite{gaussianshading}, the decoder fine-tuning method Stable Signature \cite{stable_signature}, and the backdoor-based method SleeperMark \cite{sleepermark}.

\subsection{Performance Evaluation}
We randomly selected 20 carrier images from the dataset to construct a verification set consisting of 20 tuples: the image, the watermark message, and the optimized noise. We conducted cross-verification experiments across five mainstream latent diffusion models.

\begin{figure}[t]
\centering
{\includegraphics[width=0.92\linewidth]{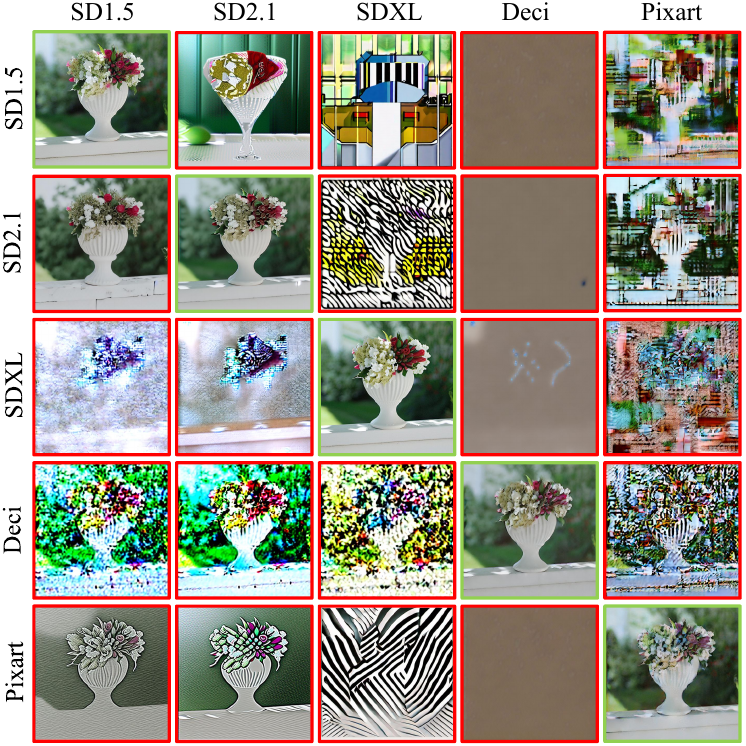}}
\caption{Visual results of cross-model reconstruction. The target model successfully recovers the watermarked anchor using the fingerprint noise, while other models generate disparate content.}
\label{experiment}
\end{figure}

As shown in Table \ref{main_experiment}, TrajPrint exhibits superior performance, characterized by a distinct diagonal dominance. Specifically, along the diagonal where the verification model matches the target model, the bit accuracy for all models exceeds 0.96, with $p$-values consistently far below $10^{-3}$. This indicates that our method generates copyright evidence with extremely high statistical significance. In contrast, within the off-diagonal regions representing cross-model verification, accuracy rapidly decays to approximately 0.55, with the majority of $p$-values showing no significance ($p > 10^{-3}$). This confirms that initializing from the inverted noise $x_T$ tightly couples the fingerprint to the target model's unique manifold, effectively preventing false positives.

In comparison, the baseline optimization initialized with random noise, while achieving relatively high verification accuracy on target models, exhibits severe cross-model transferability. Watermark accuracy reaches approximately $0.7$ across most cross-model regions. Notably, due to architectural similarities, the random noise fingerprint optimized for Stable Diffusion v1.5 achieves a high accuracy of $0.765$ on v2.1. Since $p$-values remain below $10^{-3}$ in all regions, this baseline yields false positives even under a strict significance level ($\alpha = 10^{-3}$).
In contrast, without the manifold-specific anchor $x_T$, optimizing random noise tends to generate universal adversarial perturbations that transfer across models, causing high false alarm rates.
Consequently, the random noise baseline fails to provide a unique identity marker, resulting in an unacceptably high risk of false alarms. By contrast, TrajPrint effectively anchors the optimization to the target model's unique denoising manifold via the specific anchor. This design structurally prevents the generation of universal patterns, thereby serving as the key to achieving high-precision verification with minimal false positives.

In Figure \ref{experiment}, we present the visualization results of our watermark image reconstruction. It can be observed that successful semantic reconstruction is strictly confined to the diagonal. In contrast, mismatched pairs result in severe visual degradation, appearing as either chaotic noise or complete structural collapse. This visual evidence aligns perfectly with our quantitative results. It confirms that the fingerprint is strictly coupled to the target model's manifold, explicitly lacking the transferability observed in the baseline.

\subsection{Robustness Study}

\begin{figure}[t]
\centering
{\includegraphics[width=1.0\linewidth]{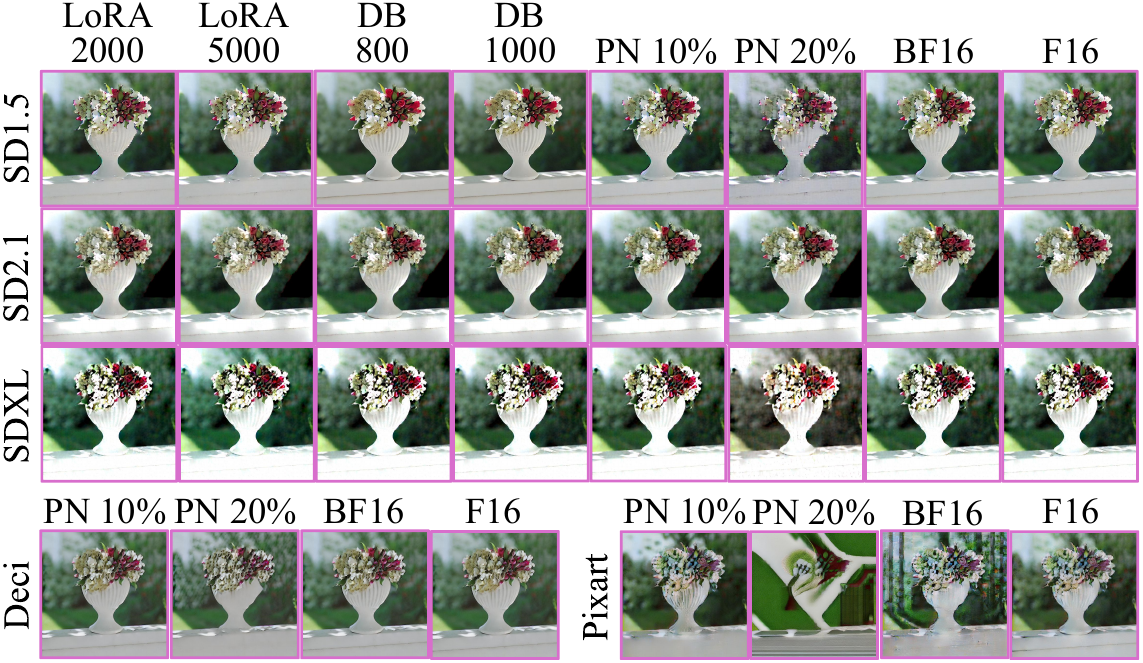}}
\caption{Visual results of the robustness study. Our method demonstrates strong stability, recovering the watermarked anchor successfully across different attack settings.}
\label{robuss_image}
\end{figure}

\textbf{Robustness to Model Modifications.}
To evaluate the robustness of our fingerprints, we simulate three categories of model post-processing operations potentially employed by attackers: parameter fine-tuning for Stable Diffusion models, including LoRA and DreamBooth (DB), unstructured magnitude pruning across the entire model (at 5\% and 10\% sparsity), and model quantization (FP16 and BFloat16).
As presented in Table \ref{robuss}, our method demonstrates exceptional robustness against all attack categories. Specifically, regarding parameter fine-tuning, our fingerprints maintain high stability even though LoRA and DreamBooth significantly alter the generative style. Even under high-intensity attacks where LoRA iteration steps are increased to 5000, the bit accuracy for SD-series models remains elevated, ranging between 0.916 and 0.959. This robustness stems from our dual-end trajectory anchoring. By locking the fingerprint to the model's fundamental denoising path rather than fragile surface features, verification remains reliable even when the generative style is altered.
Regarding model quantization, the fingerprints exhibit remarkable tolerance to numerical precision. The BA under FP16 and BFloat16 consistently exceeds 0.95, with some models approaching perfection ($>$99\%). Finally, facing highly destructive model pruning, the BA exceeds 0.91 at a 5\% pruning rate. Even when a high pruning rate of 10\% impairs model performance, the lowest BA still maintains an accuracy above 0.8. In Figure \ref{robuss_image}, we visualize the reconstruction results of our method under various attacks. As observed, the reconstructed images retain high fidelity, ensuring reliable watermark extraction.

\begin{table}[t]
    \centering
    \setlength{\tabcolsep}{2.0mm}     
    \caption{Robustness evaluation against different modification attacks on the target model.}
    \label{robuss}
    \resizebox{0.95\columnwidth}{!}{
    \begin{tabular}{c @{\hspace{0.9mm}} l c c c c c}
        \hline\hline
        \multicolumn{2}{c}{Settings} & SD1.5 & SD2.1 & SDXL & Deci & Pixart \\
        \hline
        
        \multirow{3}{*}{\shortstack{LoRA\\(steps)}}
          & 1000 & 0.938 & 0.956 & 0.959 & - & - \\
          & 2000 & 0.953 & 0.952 & 0.939 & - & - \\
          & 5000 & 0.955 & 0.951 & 0.916 & - & - \\
        \hline
        
        \multirow{3}{*}{\shortstack{Dream-\\booth\\(steps)}}
          & 500  & 0.958 & 0.936 & 0.956 & - & - \\
          & 800  & 0.935 & 0.938 & 0.936 & - & - \\
          & 1000 & 0.917 & 0.951 & 0.937 & - & - \\
        \hline
        
        \multirow{2}{*}{\shortstack{Pruning}}
          & 10\% & 0.919 & 0.972 & 0.912 & 0.969 & 0.938 \\
          & 20\% & 0.855 & 0.853 & 0.813 & 0.862 & 0.705 \\
        \hline
        
        \multirow{2}{*}{\shortstack{Quanti-\\zation}}
          & BF16       & 0.997 & 0.980 & 0.956 & 0.976 & 0.898 \\
          & FP16       & 0.998 & 0.980 & 0.977 & 0.978 & 0.953 \\
        \hline\hline
    \end{tabular}
    }
\end{table}

\begin{table}[t]
    \centering
    \setlength{\tabcolsep}{1.5mm}   

    \newcommand{\yes}{\textcolor{green!60!black}{\ding{51}}} %
    \newcommand{\no}{\textcolor{red}{\ding{55}}}             %
    
    \caption{Comparison of defense capabilities and deployment attributes with baseline methods.The symbol \yes~indicates successful defense or possession of the attribute, while \no~indicates failure.}
    \label{robuss_capability}
    \resizebox{0.95\columnwidth}{!}{
    \begin{tabular}{c l c c c c c c}
        \hline\hline
        \multicolumn{2}{c}{\makecell{Settings}} & \makecell{Rand.\\Noise} & \makecell{Gauss.\\Shad.} & \makecell{Stable\\Sig.} & \makecell{Sleeper\\mark} & \makecell{Finger\\Inv} & \makecell{Ours} \\
        \hline
        
        \multirow{3}{*}{\rotatebox{90}{LoRA}} 
          & SD1.5 & \no & \yes & \yes & \yes & \yes & \yes \\
          & SD2.1 & \no & \yes & \yes & \yes & \yes & \yes \\
          & SDXL  & \no & \yes & \yes & \yes & \yes & \yes \\
        \hline
        
        \multirow{3}{*}{\rotatebox{90}{\shortstack{Dream-\\booth}}}
          & SD1.5 & \no & \yes & \yes & \yes & \yes & \yes \\
          & SD2.1 & \no & \yes & \yes & \yes & \yes & \yes \\
          & SDXL  & \no & \yes & \yes & \yes & \yes & \yes \\
        \hline
        
        \multirow{5}{*}{\rotatebox{90}{Pruning}} 
          & SD1.5  & \yes & \yes & \yes & \yes & \yes & \yes \\
          & SD2.1  & \yes & \yes & \yes & \yes & \yes & \yes \\
          & SDXL   & \no & \yes & \yes & \yes & \yes & \yes \\
          & Deci   & \no & \yes & \yes & \yes & \yes & \yes \\
          & Pixart & \yes & \yes & \yes & \yes & \yes & \yes \\
        \hline
        
        \multirow{5}{*}{\rotatebox{90}{Quantization}} 
          & SD1.5  & \yes & \yes & \yes & \yes & \yes & \yes \\
          & SD2.1  & \yes & \yes & \yes & \yes & \yes & \yes \\
          & SDXL   & \yes & \yes & \yes & \yes & \yes & \yes \\
          & Deci   & \yes & \yes & \yes & \yes & \yes & \yes \\
          & Pixart & \no & \yes & \yes & \yes & \yes & \yes \\
        \hline
        
        \multicolumn{2}{l}{Modi-free?} & \yes & \no & \no & \no & \yes & \yes \\
        \multicolumn{2}{l}{Atom Infer?}  & \yes & \yes & \yes & \yes & \no & \yes \\
        \hline\hline
    \end{tabular}
    }
\end{table}

\textbf{Defense Capability Comparison.} 
We conducted a multi-dimensional qualitative comparison between our method and five representative baseline schemes. Our method uses statistical significance ($p < 10^{-3}$) as the criterion for success, whereas baseline methods strictly adhere to the verification protocols defined in their original papers. Beyond resistance to attacks, we evaluated two critical deployment attributes: modification-free (requiring no model modification) and atomic inference (relying solely on standard input-output interfaces).
The results are summarized in Table \ref{robuss_capability}. It is evident that while simple random noise optimization offers deployment flexibility, it fails completely under LoRA and DreamBooth fine-tuning. This proves that perturbations lacking trajectory anchoring are extremely fragile. Although Stable Signature, SleeperMark, and Gaussian Shading demonstrate excellent robustness against various attacks, they mandate invasive training or parameter modification, thereby compromising the original integrity of the model. Meanwhile, FingerInv shows good verifiability against various attacks but fails to meet the requirement for atomic inference.
In contrast, our method not only maintains successful verification under all high-intensity attacks but also achieves this without model modification and supports atomic verification. This underscores its practicality and superiority in real-world environments.

\subsection{Ablation Study}

We conduct a series of ablation studies to validate the contributions of key components in our proposed TrajPrint framework and analyze its sensitivity to various hyperparameters.

\textbf{Impact of Trajectory Anchoring Constraints.} We validate the necessity of the dual-end anchoring constraints, perceptual alignment $L_{rec}$, and latent regularization $L_{reg}$, for ensuring specificity (Table \ref{ablation1}). Removing either term significantly increases cross-model accuracy. Notably, eliminating both causes non-target accuracy to surge to approximately $0.7$, indicating the emergence of transferable universal adversarial perturbations. This confirms that both $L_{rec}$ and $L_{reg}$ are essential for anchoring the fingerprint to the model's intrinsic manifold and preventing false positives.

\begin{figure}[t]
\centering
\includegraphics[width=1.0\linewidth]{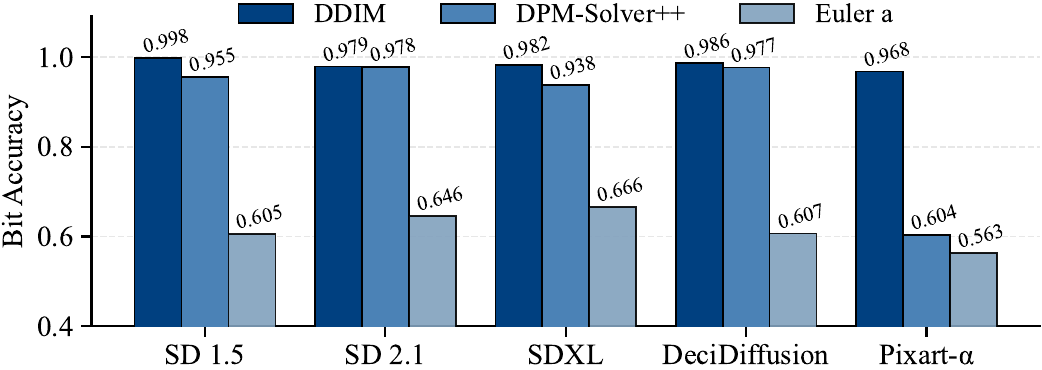}
\caption{Impact of sampling schedulers on verification. Deterministic samplers (DDIM, DPM++) maintain high bit accuracy, whereas stochastic sampling (Euler a) causes marked degradation.}
\label{ablation2}
\end{figure}

\begin{figure}[t]
\centering
\includegraphics[width=1.0\linewidth]{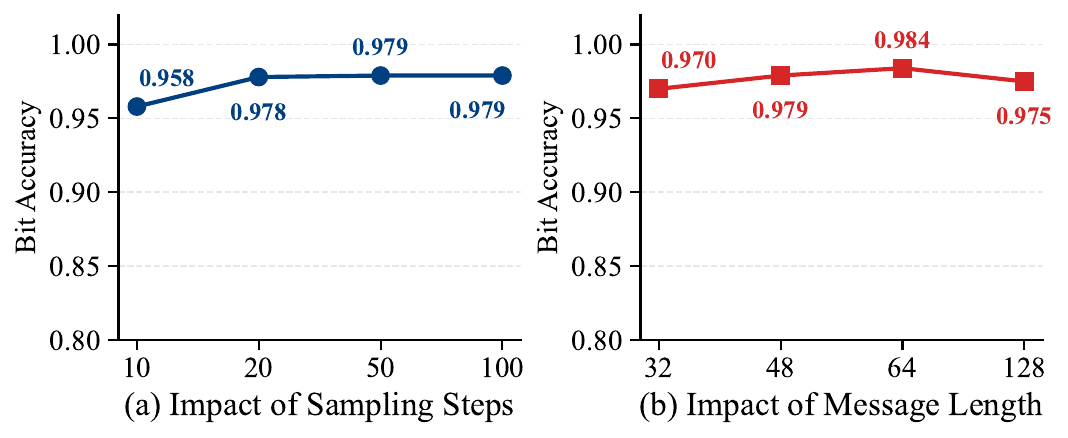}
\caption{Sensitivity analysis of (a) the number of sampling steps and (b) the watermark message length.}
\label{ablation3}
\end{figure}

\textbf{Impact of Inference Configuration.} We evaluate the fingerprint's robustness to different inference settings during verification. First, regarding sampling schedulers (Figure \ref{ablation2}), the fingerprint maintains high bit accuracy with deterministic samplers (DDIM, DPM-Solver++) but degrades with the stochastic sampler Euler a. This corroborates that the fingerprint relies on the deterministic ODE trajectory, while stochastic noise injection disrupts this locked path. Second, regarding sampling steps (Figure \ref{ablation3} (a)), accuracy remains stable ($>$0.97) even when steps are reduced from 100 to 20. This robustness indicates the fingerprint is tied to the overall trajectory, not its fine-grained discretization, ensuring practical efficiency.

\begin{table}[t]
    \centering
    \renewcommand{\arraystretch}{1.05} 
    \caption{Ablation study on the trajectory anchoring constraints. We report the bit accuracy on different target model. The result demonstrates the necessity of full constraints for model specificity.}
    \label{ablation1}
    \resizebox{0.90\columnwidth}{!}{
    \begin{tabular}{l c c c c c}
        \hline\hline
        Settings & SD1.5 & SD2.1 & SDXL & Deci & Pixart \\
        \hline
        w/o Either & 0.999 & 0.729 & 0.667 & 0.642 & 0.709 \\
        
        \hline
        w/o $L_{reg}$ & 0.999 & 0.692 & 0.646 & 0.655 & 0.701 \\
        \hline
        w/o $L_{rec}$ & 0.998 & 0.688 & 0.631 & 0.632 & 0.646 \\
        \hline
        Ours  & 0.998 & 0.582 & 0.542 & 0.551 & 0.531 \\
        \hline\hline
    \end{tabular}
    }
\end{table}

\textbf{Impact of Watermark Capacity.} Finally, we assess the information payload capacity of our method. In Figure \ref{ablation3} (b), we vary the watermark message length from 32 to 128 bits. The results show that the bit accuracy remains consistently high, staying above 0.975 even with a 128-bit payload. This demonstrates that the high-dimensional latent space provides ample capacity for embedding rich copyright information without causing signal congestion, and our optimization process effectively leverages this capacity.

\section{Conclusion and Limitation}
In this paper, we propose a black-box model copyright protection framework based on denoising trajectory anchoring. Using joint optimization, we embed invisible watermarks into the initial noise, thereby transforming copyright verification into a watermark recovery task. This process requires neither model weight modification nor intervention in the sampling procedure. Experimental results demonstrate that our modification-free method, which supports atomic inference, exhibits both exceptional specificity and robustness, withstanding common model modification attacks.
A limitation lies in the requirement for the target environment to support custom initial noise injection (e.g., ComfyUI or Stable Diffusion WebUI),  limiting its applicability in API scenarios that do not accept initial noise as input.


\bibliography{anonymous}

\end{document}